\documentclass[preprint]{elsarticle}

\usepackage{array,xspace,multirow,hhline,tikz,colortbl,tabularx,booktabs,fixltx2e,amsmath,amssymb,amsfonts,amsthm}
\usepackage[ruled,linesnumbered, noend]{algorithm2e}
\usepackage{verbatim,ifthen}

\usepackage{url}
\usepackage{enumerate}
\usepackage{pifont}
\usepackage{calrsfs,mathrsfs}
\usepackage{lscape}
\usepackage{enumitem}
\usetikzlibrary{arrows}
\usetikzlibrary{positioning}
\usepackage{mathtools}

\usepackage{lscape}

\usepackage{amssymb}
\usepackage{pifont}

	\newtheorem{definition}{Definition}%
	\newtheorem{example}{Example}

\definecolor{light-gray}{gray}{0.95}
	\newcommand{\pref}{\succ \xspace}

	\newcommand{\reals}{\mathbb{R}}

	\usepackage{enumitem}
	\setenumerate[1]{label=\rm(\it{\roman{*}}\rm),ref=({\it\roman{*}}),leftmargin=*}

	\newlength{\wordlength}

	\newcommand{\set}[1]{\{#1\}}

	\newcommand{\calX}{\mathcal{X}}
	\newcommand{\calA}{\mathcal{A}}
	
	\newcommand{\UW}[1][]{\ensuremath{\ifthenelse{\equal{#1}{}}{\mathit{UW}}{\mathit{UW}(#1)}}}
	\newcommand{\EW}[1][]{\ensuremath{\ifthenelse{\equal{#1}{}}{\mathit{EW}}{\mathit{EW}(#1)}}}
	\newcommand{\NW}[1][]{\ensuremath{\ifthenelse{\equal{#1}{}}{\mathit{NW}}{\mathit{NW}(#1)}}}

	\usepackage{boxedminipage}
	\usepackage{xspace}

	\newcommand{\naturals}{\ensuremath{\mathbb{N}}}

\begin{document}

\title{Participatory Budgeting:\\ Models and Approaches}

	%
	%
	
		%
		\author{Haris Aziz}\ead{haris.aziz@unsw.edu.au}
		\address{UNSW Sydney and Data61 CSIRO, Sydney, Australia}
		\author{Nisarg Shah} \ead{nisarg@cs.toronto.edu}
		\address{University of Toronto, Canada}




\begin{abstract}
{Participatory budgeting is a democratic approach to deciding the funding of public projects, which has been adopted in many cities across the world. We present a survey of research on participatory budgeting emerging from the computational social choice literature, which draws ideas from computer science and microeconomic theory. We present a mathematical model for participatory budgeting, which charts existing models across different axes including whether the projects are treated as ``divisible'' or ``indivisible'' and whether there are funding limits on individual projects. We then survey various approaches and methods from the literature, giving special emphasis on issues of preference elicitation, welfare objectives, fairness axioms, and voter incentives. Finally, we discuss several directions in which research on participatory budgeting can be extended in the future.}
\end{abstract}

	%

\maketitle

\section{Introduction}

Participatory budgeting (PB) is a paradigm which empowers residents to directly decide how a portion of the public budget is spent. Specifically, residents deliberate over spending priorities and vote over how the budget should be allocated to different public projects. Projects which receive broad support from the community are then funded through the process. 

This process was initiated as a radical democratic project in the city of Porto Alegre, Brazil, led by the leftist Workers' Party~\citep{Caba04a}. Over the next few decades, it quickly spread across the world; it has been implemented by over 1,500 municipalities~\citep{rocke2014framing} and in locations as diverse as Guatemala, Peru, Romania and South Africa~\citep{Shah07a}. The nonprofit organization \emph{Participatory Budgeting Project}\footnote{\url{http://www.participatorybudgeting.org}} alone has helped allocate more than US\$300 million of public budget in 29 cities across the US and Canada.
In Europe, the effort is led by Paris and Madrid, each spending at least \$100 million a year on public projects through participatory budgeting~\citep{pbParis,pbMadrid}. In particular, Madrid developed an online open-source platform, \emph{Decide Madrid}, which has been used by more than 30 local governments~\citep{decideMadrid}. {PB is still spreading in more regions across the world. For example, Toronto recently completed its three-year pilot of PB~\citep{TorontoPB}, and the state of New South Wales in Australia recently started PB under the name \emph{My Community Project}.}\footnote{\url{https://www.nsw.gov.au/improving-nsw/projects-and-initiatives/my-community-project/}}


Participatory budgeting is typically a long process; in many municipalities, one PB cycle takes one full year. While the exact implementation details differ from one PB instance to another, at a high level the process is composed of the following stages, which allow residents to come up with effective project proposals and provide their preferences over budget allocation~\citep{Shah07a}. 
\begin{itemize}
	\item First, the geographical region may be divided into several subregions (e.g. districts), and one PB process may be conducted in each district separately. The goal of this stage is to allow residents to focus on the projects in their own neighborhood and community. The total available budget for each district is also typically decided at this stage. 
	\item Next, residents share and discuss ideas through neighborhood meetings and online tools. This allows them to come up with preliminary project proposals. 
	\item These initial proposals are then developed into feasible proposals by focus groups and vetted by experts. Often, this is the stage where a proposal may be broken into distinct stages of implementation, and a cost estimate may be derived for each implementation stage. Projects may also be classified into categories such as recreation, infrastructure, health, education, transportation, etc. 
	\item This may be followed by several rounds of deliberation through which residents finalize a small number of proposals to be included in the final vote. 
	\item The final stage of the PB process is the voting stage. In this stage, eligible residents vote over how the public budget should be spent across the finalized proposals, and these votes are aggregated into a final budget allocation. 
\end{itemize}

Additional effort is often required to make PB a success. For example, advertisement and promotion through various channels, including social media, can be the key to increasing civic engagement. This can be crucial in encouraging various minorities to participate and ensuring that their preferences are incorporated into the decision-making. See the edited volumne by \citet{Shah07a} for a detailed survey of the entire PB process. 

In this chapter, we limit our attention to the final \emph{voting} stage of the process. That is, we assume that the project proposals have been detailed and finalized, and projects to be included on the final ballot have been filtered. We also assume that project costs have been estimated and the total available budget is known. {Research on the voting stage of the PB process focuses on four important considerations:}
\begin{enumerate}
	\item Decision space: Is the space of possible outcomes discrete (e.g. because a project can only be ``funded'' or ``not funded'')? Or is it continuous (e.g. because a project can be allocated different amounts of funds to implement it to different degrees of effectiveness)?
	\item Preference modeling: How will residents' preferences over the projects be represented for the purpose of mathematical analysis? 
	\item Ballot design (aka preference elicitation):  It is often desirable for the modeling to allow complex preferences. However, it may be infeasible to ask residents to report such complex preferences. What should the ballot ask from residents which will serve as a proxy for their preferences {and allow them to effectively convey their preferences}?
	\item Vote aggregation: How will the votes cast by residents be aggregated into a final allocation of the available public budget? {How should this process fairly incorporate aggregation the preferences of different communities of residents and efficiently allocate the public budget?}
\end{enumerate}

{The need for systematic modeling and study of these considerations has recently inspired a body of research in the computational social choice literature~\cite{BCE+14a,Endr17a}. This body of research sits at the intersection of computer science, social science, and economics, and aims to use mathematical modeling and algorithmic techniques to design different approaches to the voting stage of the PB process.}\footnote{In several PB programs, the final decision of budgets is entirely done by deliberation by a small subset of residents. Deliberation has its advantages and disadvantages. While it allows an in-depth discussion of possible outcomes from different perspectives, it also poses the danger of the process being dominated by an unrepresentative subset of the residents, thus thwarting the democratic objective.} {The goal of this chapter is to present a comprehensive overview of the research on PB in computational social choice. Some of the original contributions of this chapter include providing a unifying theoretical framework which allows viewing the different research works through a single lens, and providing a taxonomy of the different PB models which highlights their unique modeling choices.}



\paragraph{Outline:} The organization of this chapter is as follows. In Section 2, we present a general mathematical formulation of PB, and list several prominent features that distinguish different implementations of PB. Next, we focus on popular PB models that make specific design choices in terms of these features, and present a taxonomy of these models. We pay special attention to the representation and elicitation of preferences, and popular desiderata underlying the vote aggregation methods. 

In Section 3, we survey the research on the ``integral model'' of PB, in which each project can only be fully funded or not funded. In Section 4, we survey the research on the ``continuous model'' of PB, in which the funding level of each project can lie on a continuum. We highlight the differences between motivations and results under both discrete and continuous models. 
Finally, in Section 5, we discuss possible extensions and directions for future research. 

\section{Mathematical Formulation}

We begin by reviewing various parameters which play a key role in formulating the participatory budgeting problem. Later, we consider specific choices of these parameters, which give rise to popular PB models. 
\begin{itemize}
	\item \emph{Residents:} There is a set of residents (a.k.a. agents or voters) $N=\set{1,\ldots,n}$. In some applications of PB, residents are divided across different geographical regions (e.g. districts or wards), and each region conducts a separate PB election, in which residents of that region vote. 
	\item \emph{Resources and budgets:} There is a set of $d$ resources, denoted $R$. The available budget is $\vec{B} = (B_1,\ldots,B_d)$, where $B_r$ denotes the amount of resource $r$ available. In most applications of PB, money is the only limiting resource. Thus, much of the work on PB in computational social choice has focused on the case of a single resource, in which case the available budget is denoted by the scalar $B$.
	\item \emph{Projects:} There is a set of projects $P=\{p_1,\ldots, p_m\}$. 
	In some applications, projects are categorized into various domains (e.g. infrastructure, security, education, health \& wellness, etc).
	\item \emph{Degree of completion:} Some projects can be completed to different degrees. For example, completing $65\%$ of a project proposing neighborhood clean-up may mean a part of the neighborhood being cleaned up. Some projects may have discrete milestones, and only the first few milestones may be achieved; e.g., a project proposing the creation of a public park may include construction of the park as the first milestone and construction of a children's playground within the park as the second milestone. For other projects, the degree of completion may be binary; they must be either fully implemented or not implemented at all {(e.g. installing a fountain in a park)}. For project $p$, let $\calX_p$ denote the set of its possible degrees of completion and $x_p$ denote its actual degree of completion in an outcome. In some models of PB, {which we refer to as \emph{bounded models}}, there is an upper bound (a.k.a. \emph{cap}) on the degree of completion of each project $p$, denoted by $q_p$. In \emph{unbounded models}, there are no such caps.\footnote{{Note that the cost function of a project and the total available budget may still induce an effective upper bound on its degree of completion.}} We discuss popular choices of $\calX_p$ in Section~\ref{sec:popular-models}. We assume $0 \in \calX_p$ for all projects $p$. Let $\calX = \calX_{p_1} \times \ldots \calX_{p_m}$. Various models of PB studied in the literature crucially differ in this key choice; {we elaborate on this in Section~\ref{sec:popular-models}}. 
	\item \emph{Costs:} Each project $p \in P$ has an associated cost function $c_p : \calX_p \to \reals^d$, where the vector $c_p(x_p)$ gives the amount of each resource required to complete project $p$ to degree $x_p$. We assume that $c_p(0) = \vec{0}$ {and $c_p$ is monotonically non-decreasing}.
	\item \emph{Budget Allocations:} An outcome $\vec{x} = (x_p)_{p \in P}\in \calX$ is characterized by the degree of completion of each project $p$. Note that this also specifies the amounts of different resources that will be devoted to each project. 
	The outcome is feasible if $\sum_{p \in P} c_p(x_p) \le \vec{B}$, where addition is component-wise. We refer to feasible outcomes as \emph{(budget) allocations}. Let $\calA$ denote the space of allocations. 
	\item \emph{Resident Preferences:} Each resident $i\in N$ has preferences over which projects should be implemented and to what degree. These can be represented through an ordinal preference relation $\pref_i$ or a cardinal utility function $u_i$ over the space of allocations $\calA$. {We elaborate on this later in Section~\ref{sec:pref-modeling}.}
\end{itemize}

\begin{figure}[h]
\centering
\begin{tikzpicture}[edge from parent path={(\tikzparentnode.south) .. controls +(0,-1) and +(0,1).. (\tikzchildnode.north)},level 1/.style={sibling distance=62.5mm},level 2/.style={sibling distance=25mm},align=center]
\node {Participatory Budgeting (PB)}
child { 
	node[name=comb] {Discrete PB}
	child { node {Bounded\\Discrete PB\\ (Combinatorial PB)} }
	child { node {Unbounded \\Discrete PB} }
}
child {
	node {Divisible PB}
	child { node {Bounded\\Divisible PB} } 
	child { node {Unbounded\\Divisible PB\\ (Portioning)} } 
};
\end{tikzpicture}
\caption{A taxonomy of participatory budgeting models.}
\label{fig:taxonomy}
\end{figure}
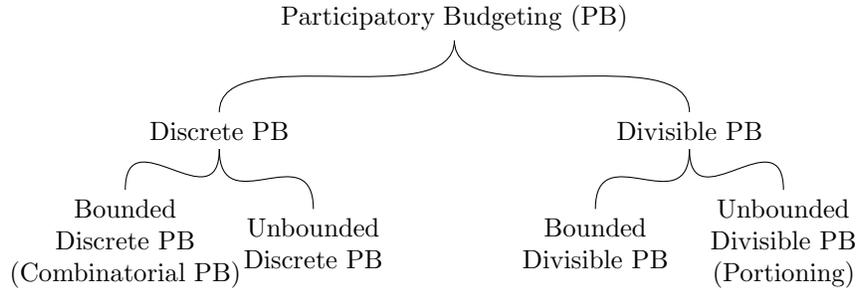

%


\subsection{Decision Space and Popular PB Models}\label{sec:popular-models}
Figure~\ref{fig:taxonomy} presents a taxonomy of the popular PB models in the literature, which crucially differ in their modeling of the possible degrees of completion of projects. We review each model in further detail. 

\paragraph{{Bounded Discrete PB (Combinatorial PB):}} This is perhaps the most widely studied and applied model of PB. In this model, projects must be either fully implemented or not implemented at all. Hence, the set of possible degrees of completion is $\calX_p = \set{0,1}$ for each project $p \in P$. {Note that this model has unit caps ($q_p = 1$).} Consequently, the cost function $c_p$ of project $p$ is effectively a vector, which indicates the amounts of different resources needed to fully implement project $p$. Feasible allocations in $\calA$ are subsets of projects which can be implemented simultaneously subject to budget constraints. In essence, this is a multi-agent variant of a multi-dimensional knapsack problem.
	

\paragraph{Discrete PB:} 
In this model, each project $p$ has discrete possible degrees of completion. However, unlike combinatorial PB, there may be more than two possible degrees of completion. These degrees of completion can also mathematically capture funding of multiple of units of the same project. For example installing 10 units of public toilets can be viewed as having a single project with 10 different degrees of completion. Typically, we assume that $\calX_p = \naturals \cup \set{0}$ for each project $p$. Note that this model has no cap on the degree of completion, but limited availability of resources may still place a natural upper bound on the degree of completion of each project in any feasible allocation. This model is applicable when each project is ambitious, and its full implementation can potentially use up all the available resources. 
	

\paragraph{Divisible PB:} In this model, it is assumed that projects can be implemented to any fractional degree. 
In the version with caps, we can assume without loss of generality that the cap is $q_p=1$, i.e., $\calX_p = [0,1]$ for each project $p \in P$; in this case, $x_p$ denotes the fraction of project $p$ that is completed. 
In practice, projects often cannot be executed to arbitrary fractional degrees. However, in cases of sufficient granularity, assuming divisible PB can help for computational reasons; more details are provided in Section~\ref{sec:divisible}.

\paragraph{Unbounded Divisible PB (Portioning):} In this model of divisible PB, there are no caps on the degrees of completion. {Thus, $\calX_p = \reals_+$ for each project $p$. When there is a single resource type (e.g. money), under mild assumptions\footnote{E.g., if the cost function of each project is strictly increasing, then there is a one-to-one correspondence between the amount of money allocated to the project and its degree of completion.}, deciding the degrees of completion of different projects is equivalent to deciding how the available budget will be divided among the projects. This setting is known as \emph{portioning} in the literature~\citep{AAC+19}.} 

\begin{example}{Example}
	Suppose there is a set of $3000$ residents $N=\set{1,\ldots,3000}$, a set of four projects $P=\set{A,B,C,D}$, and a single resource (money) with a total budget of \$$7$ million. Suppose the cost functions of the projects are as follows. 	
	
	\begin{itemize}
		\item $c_A(x_A) =$ \$$3$ million $\times  ~x_A$
		\item $c_B(x_B) =$ \$$3$ million $\times  ~x_B$
		\item $c_C(x_C) =$ \$$2$ million $\times  ~x_C$
		\item $c_D(x_D) =$ \$$2$ million $\times  ~x_D$
	\end{itemize}

	Suppose $2000$ residents like projects $A$ and $B$ equally, but derive no value from $C$ or $D$. $500$ residents derive value only from $C$, and the remaining $500$ only from $D$. 
	
	
	In the divisible PB model with unit caps {(i.e. $\calX_p = [0,1]$ for each project $p$)}, 
	we have numerous choices. For instance, we could implement $7/10$ fraction of each project. Or we could implement $1/2$ fraction of $A$ and $B$ each, and fully implement $C$ and $D$.  
	
	In the combinatorial PB model {(i.e. $\calX_p = \set{0,1}$ for each project $p$)}, given the budget of \$$7$ million, we can implement $A$ and $B$, which would make $2000$ residents very happy but $1000$ residents very unhappy, or we can implement one of $A$ and $B$ together with both $C$ and $D$, which would make $2000$ residents partially happy and $1000$ residents very happy. 
\end{example}

How do we quantify how happy the residents are? How do we make tradeoffs between such decisions? To understand this better, we need to understand modeling of resident preferences and goals of the PB process, which we review below.

\subsection{Preference Modeling and Ballot Design}

Two important decisions in the PB process are the modeling of residents' preferences (which helps in the mathematical analysis of the effectiveness of the process) and the format in which the residents cast their votes. 


\subsubsection{Preference Modeling}\label{sec:pref-modeling}
Recall that $\calA$ is the set of feasible allocations. In an expressive model, each resident $i$ has a cardinal utility function $u_i : \calA \to \reals$.\footnote{We note that in this case, much of the existing research on PB in computational social choice, inspired from classical research on voting, implicitly assumes that $u_i(\vec{0}) = 0$ and $u_i$ satisfies monotonicity (i.e. $u_i(\vec{x}) \ge u_i(\vec{x'})$ for $\vec{x} \ge \vec{x'}$). In words, implementing more projects or implementing projects to a greater degree cannot bring less happiness. In the PB context, implementing projects requires spending costly resources, which makes this assumption questionable. See Section~\ref{sec:disc} for further discussion.} If only comparisons among allocations are required, one may work instead with an ordinal preference relation $\pref_i$ over $\calA$.\footnote{The ordinal preference relation is typically assumed to be transitive and complete.} 

While such an expressive modeling allows capturing all possible preferences, it does not utilize the structure that is often present in the preferences. 
Below, we discuss three common approaches to modeling such structure. 

The first approach is to directly impose a structural assumption on the utility function or the preference relation. For example, in combinatorial PB, where allocations are effectively feasible subsets of projects and a utility function can be represented as $u_i : 2^P \to \reals$, one may impose that $u_i$ satisfies subadditivity or submodularity (when projects are substitutes of each other), or superadditivity or supermodularity (when projects are complements of each other). 

The second approach is to use spatial models, where allocations are embedded in a metric space, each resident has a preferred allocation, and her utility for another allocation is a (typically non-increasing) function of its distance to her preferred allocation. For example, \citet{GKG+19a} study $\ell_p$ normed utilities, under which the utility of resident $i$ (with preferred allocation $\vec{x}_i^*$) for allocation $\vec{x}$ is $u_i(\vec{x}) = -\|\vec{x}-\vec{x}_i^*\|_p$, where $\|\cdot\|_p$ is the $\ell_p$ norm.  

The third --- and perhaps the most popular --- approach is to model residents' preferences over individual projects and use a natural rule to extend them to preferences over allocations. Below, we review common ways to achieve this for both cardinal utility functions and ordinal preference relations. 

\paragraph{Cardinal extensions:} \citet{FGM16a} study the class of \emph{scalar separable} utility functions, whereby resident $i$ derives utility $u_{i,p} \cdot f_p(x_p)$ from each project $p$,\footnote{Note that $f_p : \calX_p \to \reals$ is a resident-independent function. Typically, it is assumed to be non-decreasing, and when $\calX_p$ is continuous, smooth and concave as well.} and her utility for an allocation $\vec{x}$ is simply additive across projects, i.e., $u_i(\vec{x}) = \sum_{p \in P} u_{i,p} \cdot f_p(x_p)$. This is a reasonable assumption when the different project proposals are completely independent of each other. In case of combinatorial PB, where $x_p \in \set{0,1}$ for each project $p$, this effectively reduces to $u_i(S) = \sum_{p \in S} u_{i,p}$ for each $S \subseteq P$; this is commonly known as an \emph{additive} utility function~\citep{BNPS17a,FMS18a,BDG18a}. A further restriction of $u_{i,p} \in \set{0,1}$ for each resident $i$ and project $p$ gives rise to \emph{dichotomous} preferences~\citep{ALT18a,FT19}, under which each resident approves or disapproves each project and only cares about the number of her approved projects that are implemented. In combinatorial PB, another common extension is the \emph{max set extension}~\citep{CNPS17}, whereby the utility of a resident for an allocation is her utility for her most favorite project that is funded: $u_i(S) = \max_{p \in S} u_{i,p}$ for each $S \subseteq P$. {This is applicable when the projects are substitutes of each other, and the resident will derive value from only one of the implemented projects.} 
	
	
\paragraph{Ordinal extensions:} When residents are assumed to have ordinal preferences over projects, we use the following notation. Each resident $i$ has a weak order preference relation $\pref_i$ over $P$. We denote by $\succ_i$ the strict part and by $\sim_i$ the indifference part of the relation $\pref_i$. We denote by $E_i^1,\dots,E_i^{k_i}$ the equivalence classes of the relation $\pref_i$ in decreasing order of preferences; thus, each set $E_i^j$ is a maximal equivalence class of objects among which agent $i$ is indifferent, and $k_i$ is the number of equivalence classes.  Given an ordinal preference relation $\pref_i$, one can induce resident $i$'s preference relation over the set of allocations $\calA$ in several natural ways. 
	
The (first order) stochastic dominance (SD) extension is defined as follows (see, e.g., \citet{brandl2016impossibility}). For allocations $\vec{x},\vec{y} \in \calA$, we say that $\vec{x} \mathrel{\pref_i^{SD}} \vec{y}$ iff $\sum_{j=1}^{l}\sum_{p\in E_i^j}x_p \geq \sum_{j=1}^{l}\sum_{p\in E_i^j}y_p \text{ for all } l\in \{1,\ldots, k_i\}.$ Since this extension requires adding the degrees of completion of different projects, it makes more sense in models where the degrees of completion of different projects are comparable (e.g. in combinatorial PB or divisible PB with unit caps). The desirable aspect of the SD extension is that $\vec{x}\mathrel{\pref_i^{SD}} \vec{y}$ is equivalent to saying that $\vec{x}$ yields at least as much utility to resident $i$ as $\vec{y}$ does, for all additive utilities over projects that are consistent with preference relation $\pref_i$. However, the SD relation is not complete (i.e. it does not compare every pair of allocations).

One popular complete extension is the lexicographic extension $\pref_i^{lex}$, under which resident $i$ is assumed to care significantly more about project $p$ than about project $p'$ whenever $p \succ_i p'$. Formally, $\vec{x} \succ_i^{lex} \vec{y}$ iff for the smallest (if any) $l$ with $\sum_{p\in E^l_i} x_p \neq \sum_{p\in E^l_i} y_p$, we have $\sum_{p\in E^l_i} x_p > \sum_{p\in E^l_i} y_p$. Once again, since this requires adding the degrees of completion of different projects, it makes more sense when the degrees of completion of different projects are comparable. However, if each resident submits a strict ordering over the projects, the equivalence classes become singletons and the definition makes sense for other models of PB too. For combinatorial PB, under the lexicographic extension, resident $i$ compares two allocations by comparing her most favorite project that implemented in each allocation, breaking ties by her next most favorite project, and so on. This is similar to the max set extension mentioned above for cardinal utility functions, and is a reasonable assumption when projects are substitutes of each other.\footnote{For instance, the projects may propose to build public parks in different locations, but a resident may only be interested in using a single park that is built closest to her home.} \citet{KPR12a} study a broader class of preference extensions that includes the lexicographic extension. 
	
Another complete extension is derived by converting ordinal preferences to cardinal preferences using \emph{scoring rules}. A scoring vector is denoted by $\vec{s} = (s_1,\ldots,s_m)$, where $s_1 \ge \ldots \ge s_m \ge 0$. Given a ranking $\succ_i$ over projects, it is assumed that resident $i$ has utility $u_{i,p} = s_k$ for project $p$, where $k$ is the rank of project $p$ under $\succ_i$. Then, any of the cardinal extensions mentioned above can be used to induce the resident's preferences over allocations. 

\subsubsection{Ballot Design} 
Even in the simplest case of combinatorial PB, there can be exponentially many allocations. This makes it infeasible to ask residents to communicate their full preferences, thus motivating the use of more restrictive preference elicitation techniques. 

For example, even when residents' preferences over allocations are induced from their ranked preferences over individual projects, asking residents to rank as many as $20$ projects\footnote{The 2018 PB in Cambridge, USA involved 20 projects: \url{https://pb.cambridgema.gov/pbcycle5}} can be tiresome, and the cognitive burden can lead to fewer residents voting or residents making poor choices~\citep{IL00}. 
Hence, most real-world applications of PB use easier ballot formats such as $k$-approval voting (where residents approve the $k$ projects they like the most), approval voting (where residents approve all projects that they like), range voting (where residents rate projects), or knapsack voting (where residents provide the ideal allocation according to their preferences)~\citep{GKS+19a}.

\subsection{Vote Aggregation}

Once the representation of residents' preferences and the format in which they cast their votes are decided, the next challenge is to decide how to aggregate their votes into a collective outcome, which can be a single allocation or a distribution over allocations (if randomization is permitted). 

This is perhaps the most well-studied aspect of the entire PB process. We review several prominent approaches to vote aggregation in computational social choice. 

\subsubsection{Welfare Maximization}

This approach is applicable when individual preferences are represented as cardinal utility functions. It uses a \emph{social welfare function}, which combines individual utility functions of residents into a societal utility function, and finds an allocation maximizing this societal utility function. There are three popular social welfare functions:

\begin{itemize}
	\item The \emph{utilitarian welfare} of an allocation is the sum of utilities it gives to residents: $\UW(\vec{x}) = \sum_{i \in N} u_i(\vec{x})$ for $\vec{x} \in \calA$.
	\item The \emph{egalitarian welfare} of an allocation is the minimum utility it gives to any resident: $\EW(\vec{x}) = \min_{i \in N} u_i(\vec{x})$ for $\vec{x} \in \calA$. Maximizing this welfare function is seen as an extreme form of one interpretation of fairness. 
	\item The \emph{Nash welfare} of an allocation is the product of utilities it gives to residents: $\NW(\vec{x}) = \prod_{i \in N} u_i(\vec{x})$ for $\vec{x} \in \calA$. Maximizing this welfare function is seen as a compromise between utilitarianism and egalitrianism.  
\end{itemize}
	
\subsubsection{The Axiomatic Approach}

This approach entails specifying intuitive normative axioms that the vote aggregation rule must satisfy and searching for rules that satisfy as many of the axioms as possible. {Many compelling axioms have been proposed for PB (see, e.g., \cite{ALT18a,FT19}); the two examples given below are selected because they apply to all the models of PB we described above.}

\begin{definition}[Exhaustiveness~\citep{ALT18a}]
	A feasible outcome (i.e. allocation) $\vec{x}$ is called exhaustive if an outcome $\vec{y}$ is not feasible whenever $y_p \ge x_p$ for all projects $p$ and a strict inequality holds for at least one project. In words, it should not be possible to implement any more projects or any projects to a greater degree with the leftover budget. A vote aggregation rule is exhaustive if it always outputs an exhaustive allocation. A similar axiom is termed budget monotonicity by \citet{FT19}.
\end{definition}


\begin{definition}[Discount Monotonicity~\citep{FT19}]
	Suppose a vote aggregation rule outputs allocation $\vec{x}$. Suppose project $p$ receives a revised cost function $c'_p$ such that $c'_p(x_p) \le c_p(x_p)$ for all $x_p \in \calX_p$, and, all else being equal, the vote aggregation rule now outputs allocation $\vec{y}$. Then, $y_p \ge x_p$. In words, if a project becomes more affordable, it should not be implemented to a lesser degree.
\end{definition}

 \citet{FT19} study a number of additional axioms that are specifically applicable to settings where residents have dichotomous preferences. 

\subsubsection{Fairness}\label{sec:fairness} 

Finally, an important consideration in democratic decision making is to fairly take into account the preferences of all residents when making the collective decision. We review one fairness axiom that is applicable in our general PB framework. For other axioms that are applicable in more specific environments (e.g. with dichotomous preferences), we refer interested readers to the work of \citet{ALT18a}.

\begin{definition}[Core~\citep{FGM16a,FMS18a}]
An allocation $\vec{x}$ is said to be in the core if no subset of residents $S \subseteq N$ can find an outcome $\vec{x'}$ that is feasible given a budget of $|S|/n \cdot \vec{B}$ such that $u_i(\vec{x'}) \ge u_i(\vec{x})$ for every resident $i \in S$ and a strict inequality holds for at least one $i \in S$. 
\end{definition}

The notion of the core is based on the idea that a group of residents $S$ collectively deserve at least $|S|/n$ fraction of the budget spent on their needs. This is formalized by requiring that they not be able to allocate this fraction of the budget in a way that they prefer more.

\section{Discrete Participatory Budgeting}\label{sec:discrete}

{We are now ready to review how these approaches have been applied to different models of PB.} Recall that in the discrete model, the decision space is discrete. This model is applicable when projects can only be executed up to discrete levels. For example, a project $p$ that proposes to build public toilets in the community can have degree of completion $x_p \in \naturals \cup \set{0}$, where $x_p$ may denote the number of public toilets that are actually built; the assumption here is that each individual toilet is either built fully or not built at all. This discrete aspect affects the modeling of residents' preferences as well as the design of ballots and vote aggregation procedures. 

This model is a natural generalization of  \emph{multi-winner voting} (alternatively known as \emph{committee selection}), which has been widely studied in social choice theory~\citep{ABC+16a,FSST17a}. In multi-winner voting, there are $m$ candidates, and $k$ of them are to be selected based on voters' preferences. This is a special case of discrete PB (more specifically, of combinatorial PB) where each candidate $p$ is a project with unit cap ($q_p = 1$) and unit cost ($c_p = 1$), and the budget limit is $k$. We first review prior literature on multi-winner voting and other similar models of decision-making, and then provide an overview of various approaches to discrete PB. 

%
%
%
%
%
%
%

\subsection{Review of the Literature on Settings Related to Discrete PB} 
The simplest special case of combinatorial PB is multi-winner voting. As explained above, this is where each project has unit cost. 
A natural goal in this setting is to maximize the utilitarian welfare. In this case, each voter approves a subset of candidates, and the goal is to select $k$ candidates with the highest number of total approvals. This can be accomplished efficiently by a greedy algorithm which selects candidates one-by-one in the decreasing order of the number of approvals they receive. However, when we replace the selection of $k$ candidates by a knapsack constraint to extend this to combinatorial PB with dichotomous preferences, the problem becomes NP-hard even for a single agent~\citep{karp1972reducibility}. The greedy algorithm is still well-defined in this setting; it selects projects one-by-one in the decreasing order of the number of approvals they receive, skipping any project if its inclusion exceeds the budget limit. This aggregation rule is often used in practice.\footnote{{See, for example, the PB process in Toronto~\citep{TorontoPB} and in Cambridge~\citep{CambridgePB}.}} 
The literature also focuses on several other objectives in multi-winner voting; an excellent overview is given by \citet{FSST17a}. 


\citet{KPR12a} also focus on committee selection under knapsack constraints. However, in their model, there is a single agent with a ranking over individual candidates (projects). They explore different ways to extend this to a ranking over sets of candidates, and then select a committee of $k$ candidates subject to a knapsack constraint. A similar model has been considered by \citet{DSW11a}.

Another model related to PB is \emph{combinatorial public projects} (CPP)~\citep{FSS08a}. In this model, a set of public projects are to be selected subject to a resource constraint, similarly to PB. However, research on CPP focuses on designing truthful mechanisms by charging payments to agents~\citep{dughmi2016optimal}, which is unrealistic in the PB setting. 

{\citet{CFS17a} propose the \emph{public decision making} model, in which there are a finite number of \emph{issues}, and for each issue, there is a set of alternatives which can be implemented. A feasible outcome consists of choosing a single alternative corresponding to each issue. One can view this as a special case of discrete PB, where each issue corresponds to a unique type of resource of which one unit is available and implementing any alternative of an issue requires the full one unit of the corresponding resource. One of the fairness definitions they propose is \emph{proportionality up to one issue}, which is a relaxation of the core defined in Section~\ref{sec:fairness}.}

\citet{lu2011budgeted} introduce \emph{budgeted social choice}. In this framework, each alternative has a cost and the goal is to select a set of alternatives subject to budget constraint. This framework generalizes the PB model we discuss as it allows the cost of an alternative to be dependent on the number of voters who derive utility from that alternative. However, they work with a limited modeling of preferences wherein there is a common \emph{positional scoring rule} which maps an alternative's rank in a voter's preference ranking to the voter's utility for the alternative. This is also a common approach in the resource allocation literature~\citep{BoLa11a}. 

\subsection{Approaches to Discrete PB}
We now provide an overview of various approaches to PB in the discrete model. 

\paragraph{Welfare maximization:} This is a natural approach when residents have cardinal utilities. Maximizing the utilitarian welfare subject to the budget constraint is the classic knapsack problem. While the problem is NP-hard~\citep{karp1972reducibility}, there exists a pseudo-polynomial time dynamic programming algorithm and a fully polynomial time approximation scheme (FPTAS)~\citep{vazirani2013approximation}. 

\citet{FST+17a} consider the combinatorial PB model and study the computational complexity of maximizing (1) the utilitarian social welfare with respect to additive utilities, (2) the utilitarian social welfare with respect to the max extension, and (3) the Nash social welfare with respect to additive utilities. All of these problems are NP-hard except under severe restrictions on residents' utility functions. 


\paragraph{Elicitation:} Eliciting cardinal utilities in the PB context can be challenging and may impose high cognitive burden on the voters, potentially even deterring them from voting in the first place~\citep{GKG+19a}. This has inspired a line of research that aims to propose simpler ballot formats. \citet{GKS+19a} introduce \emph{knapsack voting}, under which residents simply report their favorite allocation. They suggest aggregating knapsack votes using a greedy approach. In combinatorial PB, the number of ``approvals'' that a project receives is the number of residents who include the project in their favorite allocation, and projects are selected in the decreasing order of their number of approvals subject to the budget constraint. In discrete PB, the algorithm starts by setting the degree of completion of each project to zero. Then, at each step, it considers increasing the degree of completion of each project to its next possible value, and among the feasible improvements, selects the one which is part of the favorite allocation of most voters. They show that this approach has compelling welfare and incentive guarantees under restrictive models of resident preferences.

\citet{BNPS17a} compare knapsack voting to three other ballot formats: ranking projects by value, ranking projects by value-for-money, and a new format that they call \emph{threshold approval voting}. They use the \emph{implicit utilitarian voting} framework~{\citep{PR06,BCHL+15}}, in which residents are assumed to have underlying cardinal utilities and the information they provide on the ballot is treated as a proxy for their underlying utilities. They show that in the worst case, knapsack voting leads to an exponentially bad approximation of utilitarian welfare, ranking projects by value or value-for-money leads to a polynomially bad approximation, and threshold approval voting leads to a logarithmic approximation. 

\paragraph{Incentives:} Another line of research studies residents' incentive to misreport their preferences in order to induce an outcome that is better than what would be implemented had they reported their honest preferences. While all reasonable deterministic aggregation rules are susceptible to such manipulations due to the classic impossibility result by \cite{Gibb73a} and \cite{Satt75a}, \citet{BDG18a} show that when residents rank projects by value, there exist randomized aggregation rules that are \emph{strategyproof} (i.e. provide no incentive to residents to misreport) and achieve nearly optimal approximation of utilitarian welfare among all {(even non-strategyproof)} aggregation rules, implying that, {at least when randomization is allowed,} strategyproofness does not impose a significant burden in this setting. 

\paragraph{Axiomatic desiderata:} \citet{ALT18a} begin the formal study of defining axioms for PB that encode principles of proportional representation. They design algorithms for combinatorial PB with dichotomous preferences which satisfy such axioms. Their central result is an algorithm called the \emph{generalized Phragmen's sequential rule}, which computes an allocation satisfying a reasonable proportional representation property. 
{An alternative simple method, which is motivated by proportional representation concerns, is discussed by \citet{Aziz19a}.} \citet{FaTa18a}, in the same model, focus on monotonicity-style axioms of a range of aggregation rules. The common approach in this line of work is to focus on axioms and algorithms for multi-winner voting~\citep{ABC+16a,FSST17a} and extend them to combinatorial PB. 

Since combinatorial PB is more general than multi-winner voting, the negative existential and computational results from multi-winner voting carry over directly. For example, strategyproofness and any weak notion of proportional representation are inherently incompatible~\citep{Pete18a}, and finding Pareto optimal allocations is typically NP-hard~\citep{aziz2016computing}. 

\citet{condorcetbudgeting} generalize Condorcet's principle (also referred to as popularity in allocation settings) to PB and devise an algorithm to compute allocations satisfying this principle. Since Condorcet's principle is based on majoritarian comparisons, the approach is not well-suited for proportional representation of minorities, which is a prime concern in real-world applications of PB~\citep{inclusivePB,inclusivePB2}. 

\paragraph{Fairness:} \cite{FMS18a} investigate fairness in a setting that generalizes discrete PB. In particular, they show that allocations in the core may not exist in discrete PB with additive utilities, but allocations that provide a logarithmic approximation of the core are guaranteed to exist and can be computed efficiently. {For the special case of discrete PB with binary additive utilities (where each resident has utility $1$ or $0$ for each project, and utilities are additive across projects), it is an open question whether allocations in the core always exist.} Other work on proportional representation for the combinatorial PB model (see e.g. \citet{ALT18a}) can also be viewed as focussing on fairness. 

\paragraph{Other approaches:} \cite{RiIn08a} study the combinatorial PB model with additive utilities. They propose to use a generalization of the Kalai-Smorodinsky~\citeyearpar{KS75} cooperative game solution to find a desirable allocation. A similar approach is taken by \cite{RIFR05}.

\section{Divisible Participatory Budgeting}\label{sec:divisible}

The divisible model of PB applies to scenarios in which funding for a project can lie on a continuum. For instance, recall the example from Section~\ref{sec:discrete}, which involved a project proposing to build five public toilets in the neighborhood. We argued that this falls under the discrete model of PB, as one can only provide funding to build an integral number of toilets. In contrast, a project on maintaining the cleanliness of the toilets may be funded partially without any integral restrictions, and thus may fall under the divisible model of PB. 


While a divisible model can often be approximated by a discrete model, the divisible model is interesting in its own right as it often leads to better existential and computational results (see, e.g., the work of \cite{FMS18a}). 

\subsection{Review of the Literature on Settings Related to Divisible PB} 

A strand of the literature focuses on a multi-dimensional continuous space, where the dimensions correspond to categories of projects such as defense, health and education~\citep{GKG+19a,FPPV19a}. The model is especially relevant when deciding the composition of funding across different categories is more important than deciding particular projects within a category. This literature typically works with the spatial model of resident preferences, in which each resident has an ideal point in the space. The is closely related to spatial voting models studied in political science~\citep{spatial1,spatial2} 

Recall that in bounded divisible PB, each project has a cap of $q_p = 1$ (without loss of generality). One might consider a special case of this setting, where the cost function of each project $p$ is given by $c_p(x_p) = x_p$, and the total budget is $B=1$. Thus, the set of feasible allocations is given by $\set{\vec{x} : \sum_{p \in P} x_p \le 1}$. This can be thought of as portioning (also called fair mixing)~\citep{ABM19}, where $x_p$ can be thought of as the fraction of budget devoted to project $p$ (after appropriate renormalization of cost functions). As noted in Section~\ref{sec:popular-models}, this is equivalent to the unbounded divisible PB model. Alternatively, one can also think of this model as representing probabilistic voting~\citep{Gibb77a}, where each project $p$ is a candidate and $x_p$ denotes the probability of choosing candidate $p$ as the winner of the election.

\subsection{Approaches to Divisible PB}

\paragraph{Welfare maximization:} \citet{GKS+19a} study the divisible model of PB with unit caps. They consider cardinal additive utilities as well as a spatial model of utilities. They point out that a simple greedy algorithm finds an allocation maximizing the utilitarian welfare. This algorithm sorts the projects in decreasing order of their value-for-money ($\sum_{i\in N} u_i(p) / c_p$), and fully funds projects in this order until no more projects can be fully funded. The remaining budget is allocated to partially fund the next project in the list. 

\citet{GKG+19a} consider a model of decision-making in continuous spaces that is more general than PB. In their model, each resident $i$ has a favorite point $\vec{x_i}$ in the space, and her utility for an allocation $\vec{x}$ is based on the $\ell_p$ norm, specifically, $u_i(\vec{x}) = -\|\vec{x}-\vec{x_i}\|_p$. They study a class of algorithms called iterative local voting (ILV), in which residents are iteratively asked to modify the current allocation by moving it towards their favorite allocation within a local neighborhood until convergence. This approach adopts the classic stochastic gradient descent (SGD) method from optimization to multi-agent decision-making. 

\paragraph{Fairness:} \citet{fain2016core} study proportional representation in the divisible model of PB with scalar separable utility functions. 
They argue that a variant of the classic game-theoretic notion of core captures fairness in the setting. Recall that an allocation $\vec{x}$ is in the core if \emph{no} subset of voters $S$ can use $|S|/n$ of the budget to find an allocation which is no less appealing than $\vec{x}$ to any member of $S$ and strictly more appealing than $\vec{x}$ to some member of $S$. They show that in the divisible model, there always exists an allocation in the core, and present a polynomial-time algorithm for computing one. 


Recall that in portioning (i.e. unbounded divisible PB), there are no caps on the degrees of completion. Hence, as discussed in Section~\ref{sec:popular-models}, one can normalize the model such that feasible allocations satisfy $\sum_{p \in P} x_p \le 1$, and $x_p$ can be thought of as the fraction of budget devoted to project $p$~\citep{AAC+19}. \citet{ABM19} consider portioning --- they refer to it as fair mixing --- with dichotomous preferences, 
which was originally studied by \citet{BMS05a}. They consider the relative merits of several rules. They show that maximizing Nash welfare satisfies several strong fairness properties including the fact that it finds an allocation in the core. It also satisfies the \emph{average fair share} (AFS) guarantee, which requires that for any set of residents $S$, the average utility of residents in $S$ is at least $|S|/n$. Another rule with several desirable properties that emerges from their study is the \emph{conditional utilitarian rule} (CUT). In CUT, each resident is assumed to have a `personal budget' that is $1/n$ fraction of the total budget.\footnote{Note that CUT easily generalizes to the case where agents have unequal personal budgets.} Each resident spreads her personal budget among the projects that she likes that are liked by the most number of residents. Unlike the maximum Nash welfare solution, CUT does not satisfy Pareto optimality, AFS, or core fairness. 

\citet{AAC+19} consider the portioning problem where voters express preferences over individual projects and use the stochastic dominance (SD) relation to reason about their preferences over probability distributions over projects. They find that using a scoring rule as a proxy for utilities consistent with the ordinal preferences and then maximizing the Nash social welfare with respect to these proxy utilities achieves desirable notions of fairness. Another notable example of algorithms for portioning is the egalitarian simultaneous reservation algorithm~\citep{AzSt14a} for agents with weak and transitive ordinal preferences over projects. 


\paragraph{Incentives:} 
\citet{ABM19} prove that for the portioning problem with dichotomous preferences, both CUT and the Nash welfare maximizing rule provide strict incentives for residents to participate in the voting process. Additionally, CUT is strategyproof (i.e. provides voters no incentive to misreport), while maximizing Nash welfare is not. 

\citet{FPPV19a} consider the spatial voting model and focus on the setting in which a resident's disutility for an allocation is its $\ell_1$ distance from the resident's ideal allocation (a.k.a. bliss point). They present the independent markets mechanism, which is both strategyproof and satisfies a basic notion of proportional representation.

\section{Extensions and Future Directions}\label{sec:disc}
{We presented a survey of research on participatory budgeting. Specifically, we presented a mathematical model which classifies existing research across different axes, and surveyed  computational and axiomatic approaches proposed in the literature.}

Research on participatory budgeting within computational social choice is still in its infancy. As such, only limited models of resident preferences and decision-making processes have been explored. {Further research is required to design better theoretical models and approaches for participatory budgeting, and bridge the gap between theory and practice. To that end, it is crucial to leverage insights from various disciplines such as computer science, social science, microeconomics, and public policy.} In this section, we review several directions in which these models can (and should) be extended to bring them closer to reality and for them to inform real-world implementations of PB. 

\begin{itemize}
\item {\emph{Multi-dimensional constraints:}} Much of the literature focuses on a single, knapsack-style constraint which stems from a limit on the available money. While current real-world implementations of PB also only deal with money, this is one dimension where further research on practical approaches to conducting PB with multi-dimensional constraints that capture other types of costs (e.g. costs to the environment) can lead the frontier of novel PB implementations. 

\item {\emph{Voter or voter group entitlements:}} In the classic PB setting, giving equal consideration (or weight) to all voters is a normative desideratum. However, one can imagine a more general setting in which different voters or groups of voters bring different resources to the system, and their entitlements need to reflect their contributions. In other words, different voters or groups of voters may have a claim over different parts of the universal budget.


\item \emph{Distributional constraints:} Some applications of PB impose lower and upper bounds on the amount of funding which can be allocated to each project or each category of projects (e.g. education or healthcare). Studying the effects of such constraints on the welfare, fairness, and incentives is an interesting direction. 

\item {\emph{Hybrid models:}} While our taxonomy (and the literature) partitions PB models into discrete and divisible, there can be hybrid models allowing some projects to be funded only at discrete levels while others on a continuous scale.  

\item {\emph{Complex resident preferences:}} Most of the positive axiomatic and algorithmic results in the literature rely on stylized modeling of resident preferences. In practice, residents have complex preferences which stem from intricate synergies between different projects. An important research direction is to extend the existing results to more general classes of utility functions. For example, most utility functions considered in the literature treat projects as independent or substitutes; tackling complementarity in projects remains an interesting challenge.  

\item  {\emph{Initial endowments:}} In reality, some projects may already have some funds allocated to them (e.g. through previous iterations of PB). The choice between extending funding to existing projects versus funding new projects can give rise to novel challenges. 

\item  {\emph{Pledges:}} In some cases, private organizations may be willing to pledge financial support to a project conditional on certain projects receiving at least a minimum level of funding. This can affect the choice of vote aggregation methods. 

\item  {\emph{Market-based approaches:}} In recent years, market-based approaches have been investigated for public decision making~\citep{garg2018markets}. Extending this line of approach to PB can lead to intuitive mechanisms for PB. 
 
\item {\emph{Strategic agent models:}} The work on incentives in PB focuses on strategyproofness, which aims to prevent manipulations by rational agents. However, people often do not act like rational agents. An interesting direction is to use insights from behavioral game theory, develop models of realistic manipulations that residents may use in PB, and design algorithms to prevent such manipulations.
 
\item {\emph{The role of information:}} It requires significant effort to inform residents of the costs, benefits, and complexities of different projects. The manner in which this information is communicated can have significant effect on the preferences of the residents; this is a complex issue which requires a detailed study. 

\item {\emph{An end-to-end model:}} Finally, as we mentioned at the beginning of this chapter, we only focus on the final stage of voting within the entire PB pipeline. However, the initial stages have a direct impact on the final outcome. For example, the agenda-setting phase where projects are proposed by the residents themselves crucially affects the latter stages. Formally modeling the entire PB process and designing end-to-end solutions is a complex challenge of paramount importance. 

\end{itemize}



To conclude, participatory budgeting is an important grassroots approach to democracy. Research on models, methods, and axioms for PB will provide insights that will be valuable to both the theory and practice of PB.

\section{Acknowledgements}

Haris Aziz is supported by a UNSW Scientia Fellowship. Nisarg Shah is supported by an NSERC Discovery grant. The authors thank Aditya Ganguly, Barton Lee and Dominik Peters for very helpful feedback.


\end{document}